\documentstyle[preprint,aps,floats,epsfig]{revtex}

\tightenlines

\def\appendix{\par
  \setcounter{section}{0}
  \setcounter{subsection}{0}
  \def\thesection{Appendix \Alph{section}}
  \def\thesubsection{\Alph{section}.\arabic{subsection}}
  \def\theequation{\Alph{section}.\arabic{equation}}
  \setcounter{equation}{0}}
\newcommand{\be}{\begin{equation}}
\newcommand{\ee}{\end{equation}}
\newcommand{\bear}{\begin{eqnarray}}
\newcommand{\eear}{\end{eqnarray}}

\begin{document}
\preprint{ \vbox{ \hbox{ {\bf USM-TH-123}} }} \vskip 0.5cm
\title{On Lepton Flavor Violation in Tau Decays}

\author{
G.~Cveti\v c$^1$\footnote{cvetic@fis.utfsm.cl},~~
C.~Dib$^1$\footnote{cdib@fis.utfsm.cl},~~
C.~S.~Kim$^2$\footnote{cskim@yonsei.ac.kr},~~ and~~
J.~D.~Kim$^2$\footnote{jdkim@phya.yonsei.ac.kr}}

\address{$^1$Department of Physics,
Universidad T\'{e}cnica Federico Santa Mar\'{\i}a,
Valpara\'{\i}so, Chile\\ $^2$Department of Physics and IPAP,
Yonsei University, Seoul 120-749, Korea} \vskip -0.75cm

\maketitle
\begin{center}June 12, 2002\end{center}
\begin{abstract}
\vskip-5ex 
We study lepton flavor violation (LFV) in tau decays induced by
heavy Majorana neutrinos within two models: (I) the Standard Model
with additional right--handed heavy Majorana neutrinos, i.e., a
typical seesaw--type model; (II) the Standard Model with
left--handed and right--handed neutral singlets, which are
inspired by certain scenarios of $SO(10)$ models and heterotic
superstring models with $E_6$ symmetry. We calculate various LFV
branching ratios and a T--odd asymmetry. The seesaw Model I
predicts very small branching ratios for LFV processes in most of
the parameter space, although in a very restricted parameter
region it can reach maximal branching ratios $Br(\tau\to\mu\gamma)
{\sim} 10^{-9}$ and $Br(\tau\to 3\mu) {\sim} 10^{-10}$. In
contrast, Model II may show branching ratios $Br(\tau\to e\gamma)
{\sim} 10^{-8}$ and $Br(\tau\to 3e) \stackrel{<}{\sim} 10^{-9}$,
over a sizable region of the parameter space, large enough to be
tested by experiments in the near future.
\end{abstract}

\newpage
\parskip 12pt

\section{Introduction}

One of the many puzzles remaining in the current phenomenology of
particle physics is to understand the smallness of the masses
($\stackrel{<}{\sim} 1$ eV) of standard neutrinos
$\nu_e,\nu_{\mu}$ and $\nu_{\tau}$, compared to those of charged
leptons. If neutrinos are of a Dirac nature, nonzero masses could
be obtained in the Standard Model (SM) by introduction of
(sterile) right--handed neutrinos. On the other hand, if neutrinos
are of a Majorana nature, more appealing solutions to the small
neutrino mass problem exist. In order to avoid the explicit
breaking of the SM gauge symmetry and still obtain nonzero
Majorana mass terms (via spontaneous symmetry breaking), an
additional Higgs triplet is needed in the SM. The latter would
result in physical Goldstone bosons, but these have been excluded
by experiments at LEP. On the other hand, various models in the
context of extended gauge structures result in Majorana mass terms
and could give possible solutions to the neutrino mass problem. An
appealing solution is the seesaw mechanism~\cite{seesaw} within
the framework of $SO(10)$ or left--right symmetric models. In the
conventional seesaw models, the effective light neutrino masses
are within the scales of eV to MeV via a relation involving the
hierarchy between very large Majorana masses and Dirac masses
comparable to those of charged leptons.
Another possible solution was investigated in the framework of
heterotic superstring models~\cite{E6} with $E_6$ symmetry or
certain scenarios of $SO(10)$ models~\cite{SO10a}, where the
low--energy effective theories include new left--handed and
right--handed neutral isosinglets and assume conservation of total
lepton number in the Yukawa sector.

One possibility to test the neutrino sector lies in the study and
measurement of lepton--flavor--violating (LFV) processes, e.g.,
$\mu \to e \gamma$ or $3e$; $\tau \to \mu \gamma$ or $3 \mu$;
$\tau \to e \gamma$ or $3e$. Such processes are practically
suppressed to zero in the SM, due to the unitarity of the leptonic
analog of the CKM mixing matrix and the near masslessness of the
three neutrinos. Motivated by the aforementioned models with an
extended neutrino sector, the authors in Refs.~\cite{IP,KPS,GV}
derived analytic expressions for LFV decay rates of charged
leptons in such models with heavy Majorana neutrinos. The authors
of Ref.~\cite{OOS} gave a model--independent framework for
analyzing $\mu\rightarrow e\gamma$ and $\mu\rightarrow 3e$
processes and investigated specific features of several
Supersymmetric GUTs. They focused on Parity-- and T--violating
asymmetries involving muon polarization in the initial state.

Some generic properties of LFV processes and the corresponding
constraints on the neutrino mass matrix have been studied in
Ref.~\cite{Dib:2000ce}. Phenomenological studies of various LFV
and lepton--number--violating
processes have appeared in the literature, including direct
production of heavy Majorana neutrinos at various
colliders~\cite{CK}, heavy Majorana mediated processes
\cite{Flanz:2000ku}, and LFV processes (including
$\mu \to e \gamma$ and $\tau \to \mu \gamma$) in
supersymmetric frameworks \cite{LFVSUSY}.

In this paper we will consider LFV decays of tau leptons in the
framework of the two aforementioned models with extended neutrino
sectors. We will concentrate on the calculation of the
corresponding LFV branching ratios and the corresponding expected
numbers of events. In addition, we will calculate a T--odd
asymmetry induced by these processes.\footnote{ Under the
assumption of CPT symmetry, CP violation is equivalent to T
violation. While standard CP violation appears in the quark
sector, it could also arise, for example, in processes which
involve only elementary (SM) bosons \cite{Cvetic:1993wa} or
(heavy) Majorana neutrinos \cite{Endoh:2001hc,NP:2001}.} In
Sec.~II we review the two models in question. In Sec.~III we
present the formulas for the branching ratios of charged lepton
decays, $l\rightarrow l^{\prime}\gamma$ and
$l\rightarrow 3l^{\prime}$, and the T--odd asymmetry for
$l\rightarrow 3l^{\prime}$ within these two models. In Sec.~IV
we present approximate maximal values for LFV tau decay rates,
exploring the possibility of obtaining sizable rates that can be
measured in the foreseeable future, yet keeping consistency with
present experimental constraints. We give a summary and state our
conclusions in Sec.~V.

\section{Two neutrino--mixing models}

To set up our notation, we briefly review the two models in
question: We call Model I the SM with the addition of
right--handed neutrinos (singlets under the gauge group) and with
the seesaw mechanism involved, and Model II the SM with the
addition of both left--handed and right--handed neutral singlets.

\noindent{\bf Model I:}\ \ It is the SM with its $N_L$ standard
left--handed neutrinos ${\nu_L}_i$ and an additional set of $N_R$
right--handed neutrinos ${\nu_R}_i$, where the neutrino mass terms
(after gauge symmetry breaking), which can be written as
\begin{equation}
-{\cal L}^{\nu}_Y=\frac{1}{2}\bigg(\bar{\nu}_L,\bar{\nu}_R^c\bigg)
{\cal M} \left(
\begin{array}{c} \nu^c_L \\ \nu_R
\end{array} \right) +  {\rm h.c.} \ ,
\end{equation}
contain a $(N_L\!+\!N_R)\times(N_L\!+\!N_R)$--dimensional matrix
${\cal M}$ with a seesaw block form~\cite{seesaw}. This matrix can
always be diagonalized by means of a congruent transformation
involving a unitary matrix $U$, namely:
\begin{equation}
\label{matrixI}
{\cal M}= \left(
\begin{array}{cc}
0 & m_D \\ m^T_D & m_M
\end{array} \right) \ , \qquad
U{\cal M}U^{T} = \hat{\cal M}_d \ .
\label{congruent}
\end{equation}
The resulting $N_L\!+\!N_R$ mass eigenstates $n_i$ are Majorana
neutrinos, related to the interaction eigenstates $\nu_a$ by the
matrix $U$
\begin{equation}
\left(\begin{array}{c} \nu_L \\ \nu^c_R \end{array} \right)_a =
\sum^{N_L+N_R}_{i=1}U^{*}_{ia} ~{n_L}_i \ ,
\qquad
\left(\begin{array}{c} \nu^c_L \\ \nu_R \end{array} \right)_a =
\sum^{N_L+N_R}_{i=1}U_{ia} ~{n_R}_i .
\end{equation}
The first $N_L$ mass eigenstates are the light standard partners
of the charged leptons, while the other $N_R$ eigenstates are
heavy.
It is convenient, as done in Ref.~\cite{IP}, to introduce a
$N_L\times(N_L\!+\!N_R)$--dimensional matrix $B$ for charged
current interactions, and a $(N_L\!+\!N_R)\times
(N_L\!+\!N_R)$--dimensional matrix $C$ for neutral current
interactions
\begin{equation}
B_{l i}=U^{*}_{i l}, \hspace{1cm} C_{ij}=\sum^{N_L}_{a=1}
U_{ia}U^{*}_{ja}, \label{BC}
\end{equation}
where the charged leptons are taken in their mass basis. The ratio
between the Dirac mass ($m_D$) and the Majorana mass ($m_M$)
characterizes the strength of the heavy--to--light neutrino
mixings $(s_L^{\nu_l})^2 \equiv \sum_{h} | U_{h l} |^2$ ($\sim
|m_D|^2/|m_M|^2$), as well as the size of the physical light
neutrino masses: $m_{\nu_{light}}\!\sim\!m^2_D/m_M$. In this model
the very low experimental bounds on $m_{\nu_{light}}$
($\stackrel{<}{\sim} 1$ eV) impose severe constraints on the
$|m_D| \ll |m_M|$ hierarchy required, and consequently also on the
heavy--to--light neutrino mixings.

\noindent{\bf Model II:}\ \  It is similar to Model I, except that
it contains an equal number $N_R$ of left--handed ($S_{L i}$) and
right--handed ($\nu_{R i}$) neutral singlets~\cite{E6,SO10a}, and
the form of the mass matrix ${\cal M}$ is such that total lepton
number is conserved (although lepton flavor mixing is still
possible). After electroweak symmetry breaking, the neutrino mass
terms are
\begin{equation}
-{\cal L}^{\nu}_Y=\frac{1}{2}(\bar{\nu}_L,\bar{\nu}_R^c,\bar{S}_L)
{\cal M} \left( \begin{array}{c} \nu^c_L \\ \nu_R \\ S^c_L
\end{array} \right) + {\rm h.c.}, \quad
{\cal M}= \left(
\begin{array}{ccc}
0 & m_D & 0\\ m_D^T & 0 & m_M^T \\ 0 & m_M & 0
\end{array} \right) .
\label{M-model2}
\end{equation}
The mass matrix ${\cal M}$ is $(N_L\!+\!{\tilde N}_R)\times
(N_L\!+\!{\tilde N}_R)$--dimensional, where
${\tilde N}_R\!=\!2 N_R$ (the Dirac block $m_D$ is
$N_R \times N_L$--dimensional). When $N_R\!=\!N_L$,
this model predicts, for each of the $N_L$ generations,
a massless Weyl neutrino and two degenerate neutral
Majorana neutrinos. Consequently, the seesaw--type restriction
$m_{\nu_{light}}\!\sim\!m^2_D/m_M$ of Model I does not apply
here~\cite{BRV,Gonzalez-Garcia:1989rw}. Here it is not the
smallness of light neutrino masses but the present experimental
bounds on the heavy--to--light mixing parameters $(s_L^{\nu_l})^2
\sim |m_D|^2/|m_M|^2$ ($\stackrel{<}{\sim} 10^{-2}$, see below)
which impose a certain level of hierarchy $|m_D| < |m_M|$ between
the Dirac and Majorana mass sector. This hierarchy is in general
much weaker than in seesaw models. Although Model II features
($N_L$) massless neutrinos in the light sector, nonzero masses for
the light neutrinos can be generated by introducing small
perturbations in the lower right block of ${\cal M}$, i.e., small
Majorana mass terms for the neutral singlets $S_{L i}$, without
much effect on the mixings of heavy-to-light fields.

\section{Flavor--violating tau decays within the two models}

Recently LFV processes have been investigated extensively because
SUSY GUTs predict that the branching ratios for
$\mu\rightarrow e\gamma$
and $\mu\rightarrow 3e$ and the $\mu-e$ conversion rate in a
nucleus can reach just below present experimental
bounds~\cite{OOS,HT}. Here we address the predictions for LFV
decays of the form $l\rightarrow l^{\prime}\gamma$ and
$l\rightarrow 3l^{\prime}$ within the models of Section II.

The amplitudes for $l\rightarrow l^{\prime}\gamma$ and
$l\rightarrow 3l^{\prime}$ in terms of the model parameters were
obtained in Ref.~\cite{IP}. These processes occur only at one loop
level or higher in the (extended) electroweak theory. The
amplitude for $l\rightarrow l^{\prime}\gamma$ arises from a
$\gamma$-penguin with the photon on the mass shell and is given by
an expression of the form:
\begin{equation}
{\cal M}(l\rightarrow l^{\prime}\gamma) = i\frac{e\alpha_w}{8\pi
M^2_W}\epsilon^{\mu} G_{\gamma}^{l l^{\prime}}
\ \bar{u}_{l^{\prime}} i
\sigma_{\mu\nu}q^{\nu}(m_{l^{\prime}} P_L+m_l P_R) u_l \ ,
\label{llgamma}
\end{equation}
while the amplitude for $l\rightarrow 3l^{\prime}$ receives
contributions from $\gamma-$penguins, $Z-$penguins and box
diagrams, ${\cal M}(l\rightarrow 3l^{\prime}) = {\cal M}_\gamma +
{\cal M}_Z + {\cal M}_{Box}$:
\begin{eqnarray}
{\cal M}_{\gamma}(l\rightarrow 3l^{\prime}) &=& -i\frac{\alpha^2_w
s^2_w}{2 M^2_W}
\bar{u}_{l^{\prime}}\gamma^{\mu}v_{l^{\prime}} \
\bar{u}_{l^{\prime}}\Big[ F_{\gamma}^{l l^{\prime}}
(\gamma_{\mu}\!-\!\frac{q_{\mu} {q \llap /}}{q^2}) P_L
- G_{\gamma}^{l l^{\prime}} \ i \sigma_{\mu\nu}
\frac{q^{\nu}}{q^2}(m_{l^{\prime}}P_L\!+\!m_l P_R)\Big] u_l,
\label{gammal}
\\
{\cal M}_{Z}(l\rightarrow 3l^{\prime}) &=&
-i\frac{\alpha^2_w}{8M^2_W}\ \bar{u}_{l^{\prime}}\gamma_{\mu}P_L
u_l\ \bar{u}_{l^{\prime}}\gamma^{\mu}
\left[(2-4s^2_w)P_L-4s^2_wP_R\right]v_{l^{\prime}}
F_Z^{l l^{\prime}} \ ,
\label{Zl}
\\
{\cal M}_{Box}(l\rightarrow 3l^{\prime}) &=&
-i\frac{\alpha^2_w}{4M^2_W}\ \bar{u}_{l^{\prime}}\gamma_{\mu}P_L
u_l\ \bar{u}_{l^{\prime}}\gamma^{\mu}P_L v_{l^{\prime}}
F^{l l^{\prime}}_{Box} \ ,
\label{Boxl}
\end{eqnarray}
In the above expressions, $\epsilon^\mu$ [Eq.~(\ref{llgamma})] is
the photon polarization, $P_{R/L} = (1 \pm \gamma_5)/2$, and the
factors $G_{\gamma}^{l l^{\prime}}$, ..., $F^{l l^{\prime}}_{Box}$
are combinations of mixing matrix elements and some special
functions that appear in the loop diagrams of the corresponding
processes:
\begin{eqnarray}
G^{ll^{\prime}}_{\gamma}
&=& \sum^{N_L+{\tilde N}_R}_{i=1}B^*_{li}B_{l^{\prime}i}\
G_{\gamma}(\lambda_i) \ ,
\label{Gg}
\\
F^{ll^{\prime}}_{\gamma} &=& \sum^{N_L+{\tilde
N}_R}_{i=1}B^*_{li}B_{l^{\prime}i}\ F_{\gamma}(\lambda_i) \ ,
\label{Fg}
\\
F^{ll^{\prime}}_Z &=& \sum^{N_L+{\tilde
N}_R}_{i,j=1}B^*_{li}B_{l^{\prime}j} \Big[\delta_{ij}\
F_Z(\lambda_i)+C_{ij}\ H_Z(\lambda_i,\lambda_j) +C^*_{ij}\
G_Z(\lambda_i,\lambda_j)\Big] \ ,
\label{FZ}
\\
F^{ll^{\prime}}_{Box} &=& \sum^{N_L+{\tilde N}_R}_{i,j=1}
\Big[2B_{l^{\prime}i}B_{l^{\prime}j}B^*_{li}B^*_{l^{\prime}j}\
F_{Box}(\lambda_i,\lambda_j)
+B_{l^{\prime}i}B_{l^{\prime}i}B^*_{lj}B^*_{l^{\prime}j}\
G_{Box}(\lambda_i,\lambda_j) \Big] \ .
\label{FBox}
\end{eqnarray}
The explicit expressions for the loop functions $G_{\gamma}(x)$,
..., $F_{Box}(x,y)$ are given in Ref.~\cite{IP}, and their
arguments are $\lambda_i= m_i^2/M^2_W$, {\it i.e.} the masses
(squared) of the Majorana neutrinos inside the loop, in units of
$M_W$. Eqs.~(\ref{Gg})--(\ref{FBox}) involve a summation over all
Majorana neutrinos, ${\tilde N}_R$ being the number of heavy
ones ($=\!N_R, 2 N_R$ in Models I, II, respectively).

 From the amplitudes (\ref{llgamma})--(\ref{Boxl}), the decay rates
are obtained. Using the notation of Ref.~\cite{OOS}, they
take the form
\begin{eqnarray}
\lefteqn{
\Gamma (l \rightarrow l^{\prime}\gamma) =  \Gamma (l
\rightarrow e {\bar \nu}_e \nu_l) \cdot 384\pi^2
(1\!+\!m_{l^\prime}^2/m_l ^2)|A_R|^2 \ , }
\label{BR1}
\\
\lefteqn{
\Gamma(l\rightarrow 3l^{\prime}) = \Gamma(l \rightarrow e {\bar
\nu}_e \nu_l) \cdot \bigg\{ 2 |g_4|^2 + |g_6|^2 + 8 Re\left( e A_R
(2 g^*_4\!+\!g^*_6) \right) }
\nonumber
\\
& & + \left(1\!+\!m_{l^\prime}^2/m_l ^2\right) \left( 32
\log\left(m_l^2/3m_{l^\prime}^2\right)-104/3\right)
  |eA_R|^2
\bigg\}
\times \bigg\{ 1 + {\cal O}(m_{l^\prime}^2/m_l ^2) \bigg\} \ ,
\label{BR2}
\end{eqnarray}
where $g_4, g_6$ and $A_R$ are the coefficients of the operators
in the effective lagrangian relevant to these processes, and are
given by:
\begin{eqnarray}
g_4 &=& \frac{\alpha_w}{8\pi} \{ 2s^2_w\ F^{ll^{\prime}}_{\gamma}
+ (1-2s^2_w)\ F^{ll^{\prime}}_Z
+F^{ll^{\prime}}_{Box} \} \ ,
\label{g4}
\\
g_6 &=& \frac{\alpha_w}{8\pi}\{ 2s^2_w\ F^{ll^{\prime}}_{\gamma}
+ (-2s^2_w)\ F^{ll^{\prime}}_Z \} \ ,
\label{g6}
\\
eA_R &=& \frac{\alpha_{\rm em}}{8\pi}\ G^{ll^{\prime}}_{\gamma} \
, \label{AR}
\end{eqnarray}
where $\alpha_w\equiv g_2^2/(4 \pi) = \sqrt{2}G_FM_W^2/\pi$ and
$\alpha_{\rm em} \equiv e^2/(4 \pi)$ are the weak and
electromagnetic fine structure constants, and $s^2_w \equiv \sin
^2 \theta_w$.

A T-odd asymmetry can be defined in the decays $l\to 3 l'$ which
is sensitive to the CP phases of the neutrino mixing matrices, but
has the experimental drawback that it requires independent
knowledge of the initial lepton polarization (in this case, the
tau lepton polarization). Defining, in the CM frame, the decay
plane as the plane that contains the three final momenta,
$A_T$ is the asymmetry between the cases when the polarization of
the initial lepton points to one or to the other side of the
decay plane. Geometrically, $A_T$ is a $\phi$-angle asymmetry,
where $\phi$ is the angle between the decay plane and the plane
that contains the polarization vector of the initial lepton $l$
and the momentum of the final lepton with charge opposite to $l$
(see Ref.~\cite{OOS} for details). The explicit expression for
$A_T$ is then:
\begin{equation}
A_T = \left( \int_0 ^\pi \frac{d\Gamma}{d\phi} d\phi - \int_\pi
^{2\pi} \frac{d\Gamma}{d\phi} d\phi \right) / \Gamma
\end{equation}
and in terms of parameters (\ref{g4})--(\ref{AR}) it is
\begin{equation}
A_T =\frac{\Gamma(l \rightarrow e {\bar \nu}_e \nu_l)}
{\Gamma(l\rightarrow 3l^{\prime})} \bigg\{\frac{192}{35} Im \left(
eA_R~ g^*_4 \right)  - \frac{128}{35}Im \left( eA_R~ g^*_6 \right)
\bigg\}\times \bigg\{ 1 + {\cal O}(m_{l^\prime}/m_l ) \bigg\} ,
\label{AT}
\end{equation}

\section{Numerical Results and Discussions}

Several experiments provide constraints for the masses and mixings
of light and heavy neutrinos: Tritium beta decay provides the
present bound on the electron neutrino mass $m_{\nu_e}< 3$
eV\cite{PDG}. The solar neutrino deficit\cite{Solar} can
be interpreted either by matter enhanced neutrino oscillations if
$\Delta m^2_{sol}\!\sim\!1\times 10^{-5} \ {\rm eV}^2$ with small
or large mixing, or by vacuum oscillations if $\Delta
m^2_{sol}\!\sim\! 10^{-10}$ eV$^2$ with maximal
mixing\cite{bahcall}. Atmospheric neutrino experiments show
evidence for $\Delta m^2_{atm}\!\sim\! 2.2\times 10^{-3}$ eV$^2$
with maximal mixing\cite{atmos,bi-max}. We will assume that $\Delta
m^2_{sol}\!=\!|m_{\nu_\mu}^2\!-\!m_{\nu_e}^2|$ and $\Delta
m^2_{atm}\!=\!|m_{\nu_\tau}^2\!-\!m_{\nu_\mu}^2|$. Since $\Delta
m^2_{sol}\!<<\!\Delta m^2_{atm}$, then
$|m_{\nu_\tau}^2\!-\!m_{\nu_e}^2| \!=\! \Delta m^2_{atm}$ as well.
Since $\Delta m^2_{atm}\!<<\!3^2 \ {\rm eV}^2$, the 3 eV upper
bound applies to all three light neutrino masses: $m_{\nu_e},
m_{\nu_{\mu}}, m_{\nu_{\tau}} < 3$ eV. Experimental evidence
indicates that $\nu_{\tau}$--$\nu_e$ mixing is (nearly)
zero~\cite{KK}. Further,
Refs.~\cite{KK} investigated possible patterns of the Majorana
neutrino mass matrix which are compatible with these results and
the non-observation of neutrinoless double beta
decay\cite{faessler}. In the models we are considering, a number
of low--energy experiments set upper bounds on possible non--SM
couplings, which are characterized in Refs.~\cite{IP,LL} as
$(s^{\nu_l}_L)^2\equiv\sum_{h}|B_{lh}|^2$ (where $h$ indicates
heavy neutrinos). Recent analyses \cite{NRT}, for models where
the additional neutrinos are $SU(2)_L$--singlets, give
\begin{equation}
(s^{\nu_e}_L)^2 < 0.005,\ \ \ (s^{\nu_\mu}_L)^2 <0.002,\ \ \
(s^{\nu_\tau}_L)^2 < 0.010 . \label{s2s}
\end{equation}
There is also a theoretical constraint, a perturbative unitarity
condition (PUB) \cite{PUB1}, which states that perturbation theory
to one loop is applicable only if the decay width $\Gamma_{n_h}$
of a heavy Majorana neutrino is small compared to its mass, say,
$\Gamma_{n_h} < M_{n_h}/2$. In the limit of large masses $M_{n_h}
 >> M_W, M_Z, M_H$, the PUB constitutes an upper bound for heavy
neutrino masses:\cite{IP,Pilaftsis:1991ug,Fajfer:px}
\begin{equation}
M^2_{n_h} \sum_{l=1}^{N_L} |B_{l h}|^2 < \frac{2}{\alpha_w} M^2_W
\ , \qquad  h\!=\!1,\ldots,{\tilde N}_R. \label{PUB}
\end{equation}
In addition, there is a lower bound, $M_{n_h} > 100$ GeV, arising
from the non-observation of heavy neutrinos in experiments to
date.

As the bound on $(s^{\nu_\mu}_L)^2$ in (\ref{s2s}) is tighter than
those on $(s^{\nu_e}_L)^2$ and $(s^{\nu_\tau}_L)^2$, LFV muon
decays are more suppressed than tau decays. We will therefore
study LFV tau decays in the two models, trying  to see if
the present experimental upper bounds on $\tau \to l \gamma$ and
$\tau \to 3 l$ \cite{PDG}
\begin{eqnarray}
Br(\tau^-\rightarrow e^-\gamma) &<& 2.7\times 10^{-6} , \qquad
Br(\tau^-\rightarrow \mu^-\gamma) < 1.1\times 10^{-6} ,
\label{tlgexp}
\\
Br(\tau^-\rightarrow e^-e^-e^+) &<& 2.9\times 10^{-6}  , \quad
Br(\tau^-\rightarrow \mu^-\mu^-\mu^+) < 1.9\times 10^{-6},
\label{tlllexp}
\end{eqnarray}
can be reached theoretically once we account for all
the aforementioned constraints.

A numerical analysis of such reactions in Model I, for $N_L\!=\!3$
and $N_R\!=\!2$, has been performed in Ref.~\cite{IP}. A
comprehensive numerical analysis in Model II has been performed in
Ref.~\cite{Ilakovac:1999md}. In these two references, the analyses
were performed by starting with the neutrino eigenmasses and
specific combinations of the mixing matrix coefficients $B_{ij}$
on which restrictions were imposed. Our approach will be somewhat
different, starting with explicit mass matrices, Eqs.\
(\ref{matrixI}) and (\ref{M-model2}), and from there deriving the
masses and mixings. This approach is cumbersome if we include all
three light generations, so we will take $N_L\!=\!2$ and
$N_R\!=\!2$. In this way, $Br(\tau \to \mu \mu e, e e \mu)$
will not be considered. However, since we are particularly
interested in the largest possible branching ratios, and since
either $Br(\tau \to 3 e)$ or $Br(\tau \to 3 \mu)$ is very
suppressed (as argued below), then $Br(\tau \to \mu \mu e, e e \mu)$
is also expected to be suppressed in
comparison with the largest of the LFV branching ratios.

\subsection{Model I}
In the considered case ($N_L\!=\!2$ and $N_R\!=\!2$) for LFV
$\tau$ decays, we have two light lepton generations ($\nu_{\ell},
\ell^-$) and ($\nu_{\tau}, \tau^-$), where $\ell$ is either $e$ or
$\mu$. The structure of the CP--violating phases in these types of
models was studied in Ref.~\cite{Endoh:2001hc}. In the considered
case there are two independent physical phases ($\delta_i$,
$i\!=\!1,2$), so that the Dirac and Majorana submatrices in ${\cal
M}$ [see Eq.~(\ref{matrixI})] can be taken to be of the form
\begin{equation}
m_D=\left(
\begin{array}{cc} a & be^{i\delta_1} \\ ce^{i\delta_2} & d
\end{array}\right), \;\;\;
m_M=\left(
\begin{array}{cc} M_1 & 0 \\ 0 & M_2
\end{array}\right) ,
\label{DM}
\end{equation}
where $a,b,c,d$ are real. We take the convention $M_2 \geq M_1$.
The matrix ${\cal M}$ can be diagonalized via the congruent
transformation of Eq.~(\ref{congruent}) -- in numerical
calculations we use the diagonalization approach as
described in Ref.~\cite{AS}.

We then find the values of $m_D$ ({\it i.e.} $a,b,c,d; \delta_1,
\delta_2$) which give, for given heavy Majorana masses $M_1$ and
$M_2$, the largest possible LFV branching ratios $Br(\tau \to
\gamma \ell)$ and $Br(\tau \to 3 \ell)$.

In Model I, the transformation matrix $U$ of Eq.~(\ref{congruent})
can be presented as a product of a seesaw transformation block
matrix $U_s$ and a light-sector mixing matrix $V^{\dagger}$: $U =
V^{\dagger} U_s$. The seesaw transformation $U_s$ produces an
effective light neutrino mass matrix $m_{\nu_{light}}\approx m_D
m^{-1}_M m^T_D$ in the case $m_D\ll m_M$, namely:
\begin{equation}
m_{\nu_{light}} = \left[
\begin{array}{cc}
\left( \frac{a^2}{M_1} + \frac{b^2}{M_2} e^{2 i \delta_1} \right) &
\left( \frac{a c}{M_1} e^{i \delta_2} + \frac{b d}{M_2} e^{i \delta_1}
\right) \\
\left( \frac{a c}{M_1} e^{i \delta_2} + \frac{b d}{M_2} e^{i \delta_1}
\right) &
\left( \frac{c^2}{M_1} e^{2 i {\delta}_2} + \frac{d^2}{M_2} \right)
\end{array}
\right] \times \left( 1 + {\cal O}(m_D^2 m_M^{-2}) \right) \ ,
\label{mnulight}
\end{equation}
and LFV mixings of order $\sim\!m_D m^{-1}_M$.
The light sector mixing matrix $V^{\dagger}$, which is the
upper left part of $U$, is approximately
unitary and of the form:
\begin{equation}
V^{\dagger} \approx \left(
\begin{array}{cc}
\cos \theta & \sin \theta \exp(-i\varepsilon) \\ - \sin \theta
\exp(i\varepsilon) & \cos \theta
\end{array}
\right) \ ,
\label{mix}
\end{equation}
where $\theta=0$ and $\pi/4$ correspond to zero and maximal
mixing, respectively, and where $\varepsilon$ is a CP phase, in
general a complicated function of $\delta_1$ and $\delta_2$.

In the case $\tau\to\mu$, we indeed have maximal mixing
$\theta\to\pi/4$, according to atmospheric neutrino experiments.
We will consider first this case. If we demand that this maximal
mixing is obtained independently of the values $M_1$ and $M_2$ of
the heavy Majorana sector [see Eq.~(\ref{DM})], then the following
simple relations in the light Dirac sector are implied: $a^2 =
c^2$, $b^2 = d^2$, and $\delta_1 = \delta_2 \equiv \delta$. The
value of $\delta$ can be restricted to lie in the range $- \pi/2 <
\delta \leq \pi/2$.
The eigenmasses of the two light neutrinos are then
\begin{equation}
m_{\nu_1, \nu_2} = \left| \left[
\left( \frac{a^2}{M_1} \right)^2 +  \left( \frac{b^2}{M_2} \right)^2
+ 2 \frac{a^2}{M_1} \frac{b^2}{M_2} \cos(2 \delta) \right]^{1/2}
\pm \left( \frac{ac}{M_1} + \frac{bd}{M_2} \right) \right| \quad
\ ,
\label{mlight12}
\end{equation}
while the heavy--to--light mixing parameters (\ref{s2s})
$(s_L^{\nu_{l}})^2\equiv \sum^4_{h=3}|B_{l h}|^2$ are
\begin{equation}
(s_L^{\nu_{\mu}})^2 = (s_L^{\nu_{\tau}})^2 =
\frac{a^2}{M^2_1}  +  \frac{b^2}{M^2_2} \ (\equiv s_L^2) \ ,
\label{s2s1}
\end{equation}
and the CP--violating parameter $\varepsilon$ of Eq.~(\ref{mix}) is
\begin{equation}
\tan\varepsilon = \tan \delta \times
\frac{(a^2/M_1) - (b^2/M_2)}{(a^2/M_1) + (b^2/M_2)} \ .
\label{eps}
\end{equation}

\begin{center}
\begin{figure}[htb]
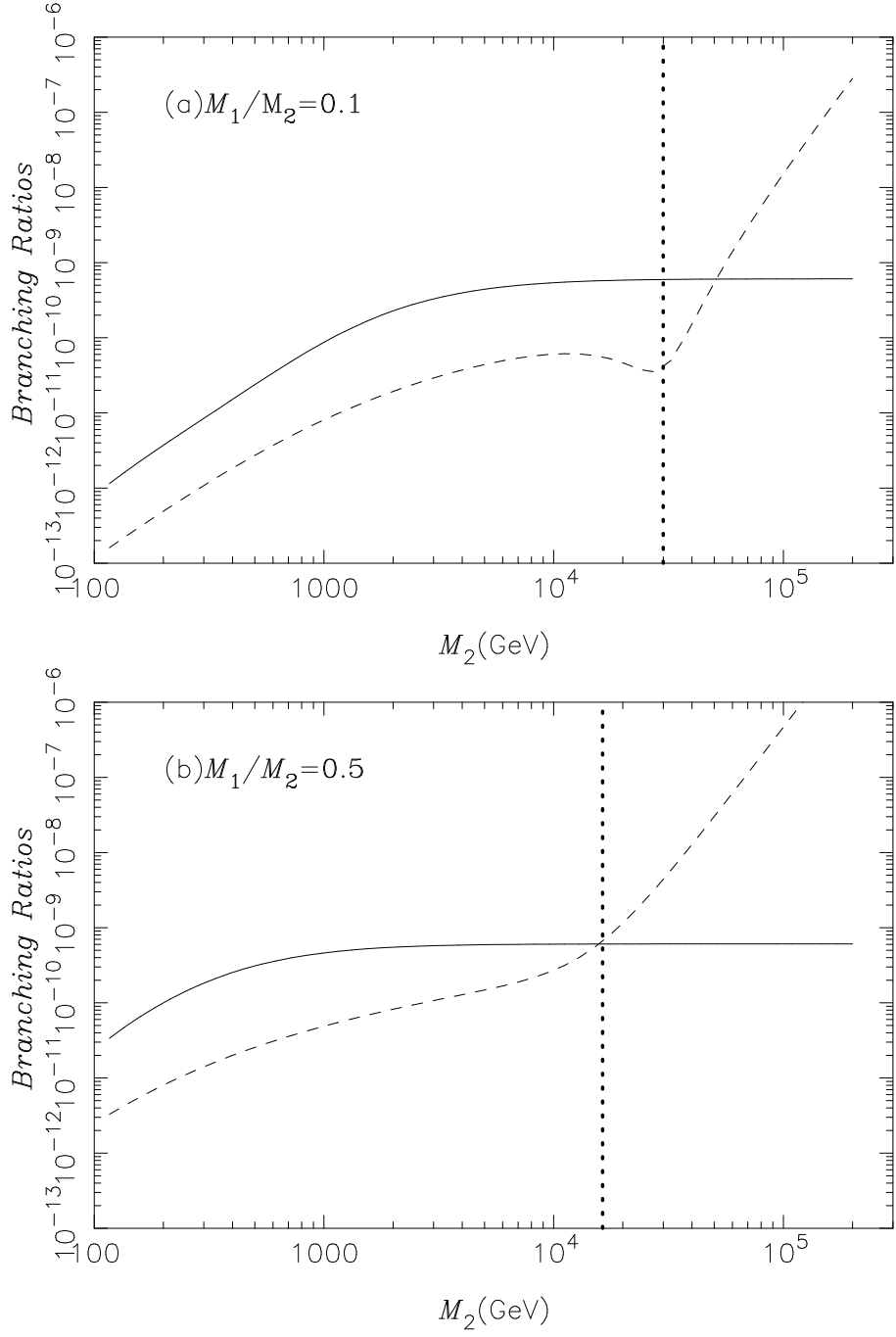

\begin{center}
\epsfig{file=cf1_01.ps,width=9cm,height=12cm,angle=-90}
\epsfig{file=cf1_05.ps,width=9cm,height=12cm,angle=-90}
\end{center}
\vskip 1.5cm
\caption{Maximal branching ratios for $\tau\rightarrow \mu\gamma$
(solid lines) and $\tau\rightarrow 3\mu$ (dashed lines) as
functions of $M_2$ in Model I, for a fixed ratio $M_1/M_2\!=\!
0.1$ (a) and $M_1/M_2\!=\! 0.5$ (b). $M_1$ and $M_2$ are
restricted to be above $100$ GeV and below the perturbative
unitarity bounds (\ref{PUB}), indicated by the vertical line. The
Dirac mass parameters are taken in the form
(\ref{abmodI})--(\ref{etamodI}) (and
$\delta_1\!=\!\delta_2\!=\!\pi/2$) which give maximal branching
ratios at any given $M_1$ and $M_2$.} \label{figBrMI}
\end{figure}
\end{center}

Now, conditions $a^2 = c^2$, $b^2 = d^2$ mean two possible cases:
(1) $a=\pm c$ and $b=\pm d$; or (2) $a=\pm c$ and $b=\mp d$.
\begin{enumerate}
\item
Case $a=\pm c$ and $b=\pm d$: then $m_{\nu_{\tau}} \geq (a^2/M_1 +
b^2/M_2)$, so $s_L^2 \leq m_{\nu_{\tau}}/M_1 < 3 {\rm eV}/M_1 \leq
3 \times 10^{-11}$. Since $Br(\tau \to \mu \gamma)$ and $Br(\tau
\to 3 \mu)$ are approximately proportional to $(s_L^{\nu_{\mu}})^2
(s_L^{\nu_{\tau}})^2$ ($\equiv {s_L}^4$), it follows that
$Br(\tau \to \mu \gamma)$ and $Br(\tau \to 3\mu)$ are below
$10^{-24}$. For muon decays, $Br(\mu \to e \gamma)$ and
$Br(\mu \to 3 e)$ are obtained by dividing the
previous values by $Br(\tau \to \mu {\overline \nu}_{\mu}
\nu_{\tau}) \approx 0.174$, thus obtaining
$Br(\mu \to e \gamma)$ and $Br(\mu \to 3 e)$ below
$10^{-23}$, values which are well below their
respective present experimental bounds ($10^{-11}$ and
$10^{-12}$).

\item
Case $a=\pm c$ and $b=\mp d$: then $m_{\nu_{\tau}} \geq 2 |
a^2/M_1 - b^2/M_2|$, with the equality being reached only when
$\delta\!=\!\pi/2$. In the latter case, $m_{\nu_{\mu}} = 0$, and
$m_{\nu_{\tau}}\!=\!2 | a^2/M_1 - b^2/M_2|\!=\!(\Delta
m^2_{atm})^{1/2} \approx 0.047$ eV. This case ($\delta\!=\!\pi/2$)
thus avoids the suppression of $s^2_L\!=\!(a^2/M_1^2 + b^2/M_2^2)$
while keeping $a^2/M_1$ extremely close to $b^2/M_2$ (a fine
tuning situation). The value of $s_L^2$ can then be saturated to
$(s_L^2)_{\rm max} = 0.002$ [Eq.~(\ref{s2s})] with the following
parameters in the Dirac matrix $m_D$:
\begin{eqnarray}
a &=& c = M_1 (s_L)_{\rm max}/\sqrt{1 + M_1/M_2} \ , \quad
b = - d = a (1 \pm \eta) \sqrt{M_2/M_1} \ ,
\label{abmodI}
\\
\eta &=& \sqrt{\Delta m^2_{atm}}
\left( 1 + \frac{M_1}{M_2} \right)
\frac{1}{4 M_1 (s^2_L)_{\rm max}}
\approx 1.17 \times 10^{-13} \times
\frac{1}{(s^2_L)_{\rm max}}
\left(1 + \frac{M_1}{M_2} \right)
\ ,
\label{etamodI}
\end{eqnarray}
and where, as mentioned, $\delta\!=\!\pi/2$. The rates $\tau \to
\mu \gamma$ and $\tau \to 3\mu$ are again practically proportional
to $(s_L^4)$ and approach their maximum (for fixed chosen values
of $M_1$ and $M_2$) in the case given by
Eqs.~(\ref{abmodI})--(\ref{etamodI}), as shown in the Appendix.
The conditions (\ref{abmodI})--(\ref{etamodI}), which are only
reached by fine tuning, give the largest possible branching ratios
in Model I, $Br(\tau \to \mu \gamma) {\sim} 10^{-9}$ and $Br(\tau
\to \mu \mu \mu) \stackrel{<}{\sim} 10^{-10}$. In
Fig.~\ref{figBrMI} we show the two branching ratios as functions
of $M_2$, for two different ratios $M_1/M_2\!=\!0.1$ and $0.5$,
accounting also for the PUB conditions (\ref{PUB}). The
CP--violating asymmetry parameter $A_T$ (\ref{AT}) is in this
case, unfortunately, equal to zero, since $\delta = \pi/2$ implies
$\varepsilon =0$ [Eq.~(\ref{eps})] and thus no CP violation. If we
move $\delta$ away from $\pi/2$, the allowed branching ratios drop
sharply, mainly due to the upper bounds $m_{\nu_{\mu}},
m_{\nu_{\tau}} < 3$ eV, i.e., a situation similar to case 1 sets
in. Accordingly, we do not consider other situations of CP
violation in Model I, as the branching ratios fall dramatically to
unobservable values outside the fine--tuning condition.

The results for $Br(\mu \to e \gamma)$ and $Br(\mu \to 3 e)$ are
again obtained by dividing the above results by $0.174$. However,
we will then obtain values above the present experimental bounds
$1.2 \times 10^{-11}$ and $1.0 \times 10^{-12}$, respectively.
We are thus led to conclude that the assumed fine--tuning
condition is not really met in this case.

\end{enumerate}

If we now consider the case $\tau \to e$, i.e.\ the processes
$\tau \to e \gamma$ and $\tau \to e e e$, the neutrino oscillation
experiments indicate that the mixing is almost zero\cite{KK}:
$\theta \approx 0$ in Eq.~(\ref{mix}). If we assume that the zero
mixing condition is fulfilled independently of the heavy Majorana
sector, we obtain the relations $ac=0$ and $bd=0$. The cases where
$a=b=0$ or $c=d=0$ give us $(s_L^{\nu_e})^2\!=\!0$ and
$(s_L^{\nu_{\tau}})^2\!=\!0$, respectively, and thus extremely
suppressed branching ratios. The cases where $a=d=0$ or $b=c=0$
give: $(s_L^{\nu_{e}})^2 (s_L^{\nu_{\tau}})^2 < m_{\nu}/M_1 <
3{\rm eV}/100{\rm GeV} = 3 \times 10^{-11}$, i.e., as in case 1
discussed previously we obtain extremely suppressed branching
ratios.

\subsection{Model II}
The neutrino mass matrix has the form of Eq.~(\ref{M-model2}).
In the considered two--generation case
($N_L\!=\!2$, ${\tilde N}_R\!=\!4$) for LFV $\tau$ decays,
the submatrices $m_D$ and $m_M$ can be taken in the form
\begin{equation}
m_D= \left(
\begin{array}{cc}
a & b e^{i\xi} \\ c e^{i\xi} & d
\end{array}
\right), \;\;\; m_M= \left(
\begin{array}{cc}
M_1 & 0 \\ 0 & M_2
\end{array}
\right) . \label{DM2}
\end{equation}
In the two--generation scheme of Model II there is
only one CP--violating phase $\xi$ \cite{BRV}.

Since the ($N_L$) light neutrinos in the model are
massless, the LFV $\tau$ decay rates will neither be affected
by the experimental light neutrino mass bounds,
nor by the solar and atmospheric neutrino experiments
and their requirements of the maximal
($\nu_{\mu}$--$\nu_{\tau}$, $\nu_e$--$\nu_{\mu}$) or minimal
($\nu_e$--$\nu_{\tau}$) mixing.\footnote{
Nonetheless, it is possible
to obtain nonzero light neutrino masses in Model II to
accommodate neutrino oscillation experiments.
Introduction of small mass terms for the neutral singlets
$S_{L i}$ gives non-zero and non-degenerate
masses of the light neutrinos, thus the possibility to
accommodate $\Delta m^2_{atm}$ without significantly
affecting the presented LFV rates.}
However, the rates will be affected by the PUB restrictions
(\ref{PUB}), as well as by the mixing parameter bounds (\ref{s2s})
as the rates are proportional to
$(s_L^{\nu_{\tau}})^2 (s_L^{\nu_{\mu}})^2$ or
$(s_L^{\nu_{\tau}})^2 (s_L^{\nu_{e}})^2$.
The $\tau\to\mu$ rates are suppressed in comparison to
$\tau \to e$ rates, because the upper
bound for $(s^{\nu_{\mu}}_L)^2$ is smaller than
that for $(s^{\nu_{e}}_L)^2$ [cf.~Eq.~(\ref{s2s})].
Therefore, we will consider $\tau \to e \gamma$
and $\tau \to 3 e$ LFV rates.

\begin{center}
\begin{figure}[htb]
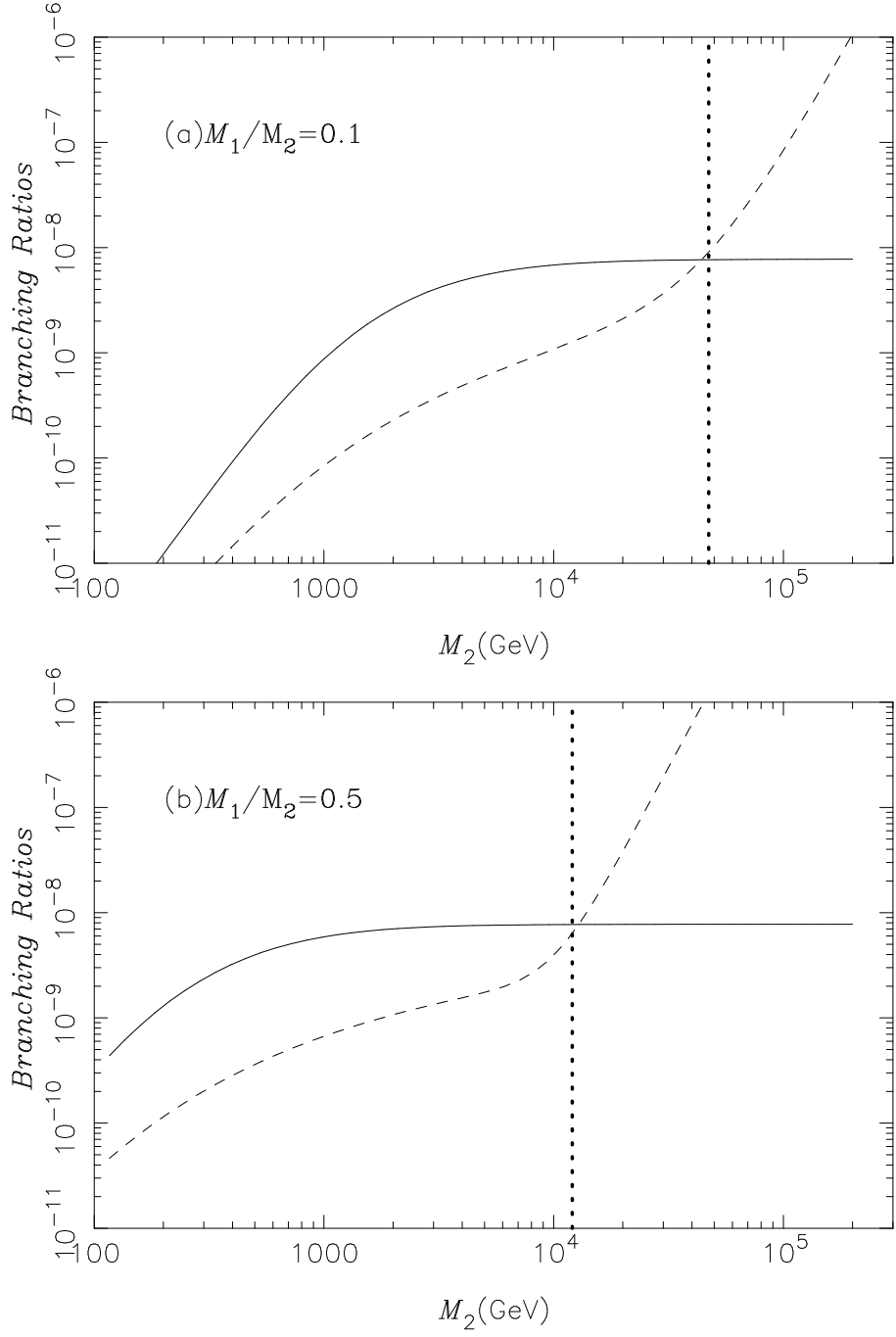

\begin{center}
\epsfig{file=cf2_01.ps,width=9cm,height=12cm,angle=-90}
\epsfig{file=cf2_05.ps,width=9cm,height=12cm,angle=-90}
\end{center}
\vskip 1.5cm
\caption{ Maximal branching ratios for $\tau\rightarrow e\gamma$
(solid line) and $\tau\rightarrow 3e$ (dashed line) as functions
of $M_2$ in Model II, for a fixed ratio $M_1/M_2\!=\! 0.1$ (a) and
$M_1/M_2\!=\!0.5$ (b). $M_1$ and $M_2$ are restricted to be above
$100$ GeV and below the perturbative unitarity bounds (\ref{PUB}),
indicated by the vertical line. The Dirac mass parameters
$a,b,c,d$ are taken in the form (\ref{amodII})--(\ref{bcdmodII})
which give approximately maximal branching ratios. Here we use the
CP phase $\xi=0$.} \label{figBrMII}
\end{figure}
\end{center}

Similarly as in Model I, we first find the Dirac parameters in
Model II such that the LFV branching ratios, at fixed Majorana
masses $M_1$ and $M_2$, are maximal. This occurs when inequality
(\ref{A2t}) in the Appendix becomes equality, and the values of
the mixing parameters $(s_L)^2$ are maximized, i.e., saturated
according to Eq.~(\ref{s2s}). In the case of no CP violation
($\xi\!=\!0$), the requirement (\ref{Schweq}) for the inequality
(\ref{A2t}) to become an equality gives the relation $a d = b c$,
while the saturation of the values of $(s_L^{\nu_{\tau}})^2$ and
$(s_L^{\nu_{e}})^2$ gives two other conditions, for the four Dirac
parameters $a, b, c, d$. This still allows us the freedom of
fixing one of the four Dirac parameters without affecting the
rates. We can, for example,
require the symmetry of the (real) $m_D$ matrix: $b\!=\!c$. All of
the above results in the following approximately ``optimized''
choice of $m_D$ parameters (when $\xi\!=\!0$):
\begin{eqnarray}
a & = & \frac{M_2}{\sqrt{(M_2/M_1)^2 + (s_{2m}/s_{1m})^2}}
\frac{s_{1m}}{\sqrt{1 - s^2_{1m} - s^2_{2m}}} \ ,
\label{amodII}
\\
b&=&c = a \times (s_{2m}/s_{1m}) \ , \quad d = a
\times (s_{2m}/s_{1m})^2 \ ,
\label{bcdmodII}
\end{eqnarray}
where $s^2_{1m} = (s_L^{\nu_{e}})^2_{\rm max} = 0.005$ and
$s^2_{2m} = (s_L^{\nu_{\tau}})^2_{\rm max} = 0.010$, according to
the bounds of Eq.~(\ref{s2s}).\footnote{ In contrast to Model I,
we do not have the requirement of
$(s_L^{\nu_{\tau}})^2\!=\!(s_L^{\nu_{\ell}})^2$ which followed
there from the maximal ($\nu_{\ell}$--$\nu_{\tau}$) mixing
condition ($\ell\!=\!\mu$ there), so that in Model II we can
saturate each of the two upper bounds (\ref{s2s}) separately.} In
Fig.~\ref{figBrMII} we present the two branching ratios $Br(\tau
\to e \gamma)$ and $Br(\tau \to 3 e)$ as functions of $M_2$, for
two fixed ratios $M_1/M_2\!=\! 0.1$ and $0.5$, and for the CP
phase $\xi =0$. We see from Fig.~\ref{figBrMII} that the LFV
branching ratios in Model II are $Br(\tau \to e \gamma)
\stackrel{<}{\sim} 10^{-8}$ and $Br(\tau \to 3e)
\stackrel{<}{\sim} 10^{-9}$. These values decrease relatively
slowly when the parameters of the Dirac sector ($a,b,c,d; \xi$)
are moved away from the ``optimal'' values. This contrasts with
Model I, where the maximal rates are reached only in a finely
tuned region of parameter space. Our maximal values of
$Br(\tau \to e \gamma)$ agree with those of Ilakovac
\cite{Ilakovac:1999md} -- ours are by about factor three
lower only because we took a different upper bound
$(s_L^{\nu_{\tau}})^2 < 0.010$ (\ref{s2s}).

\begin{center}
\begin{figure}[htb]
\begin{center}
\epsfig{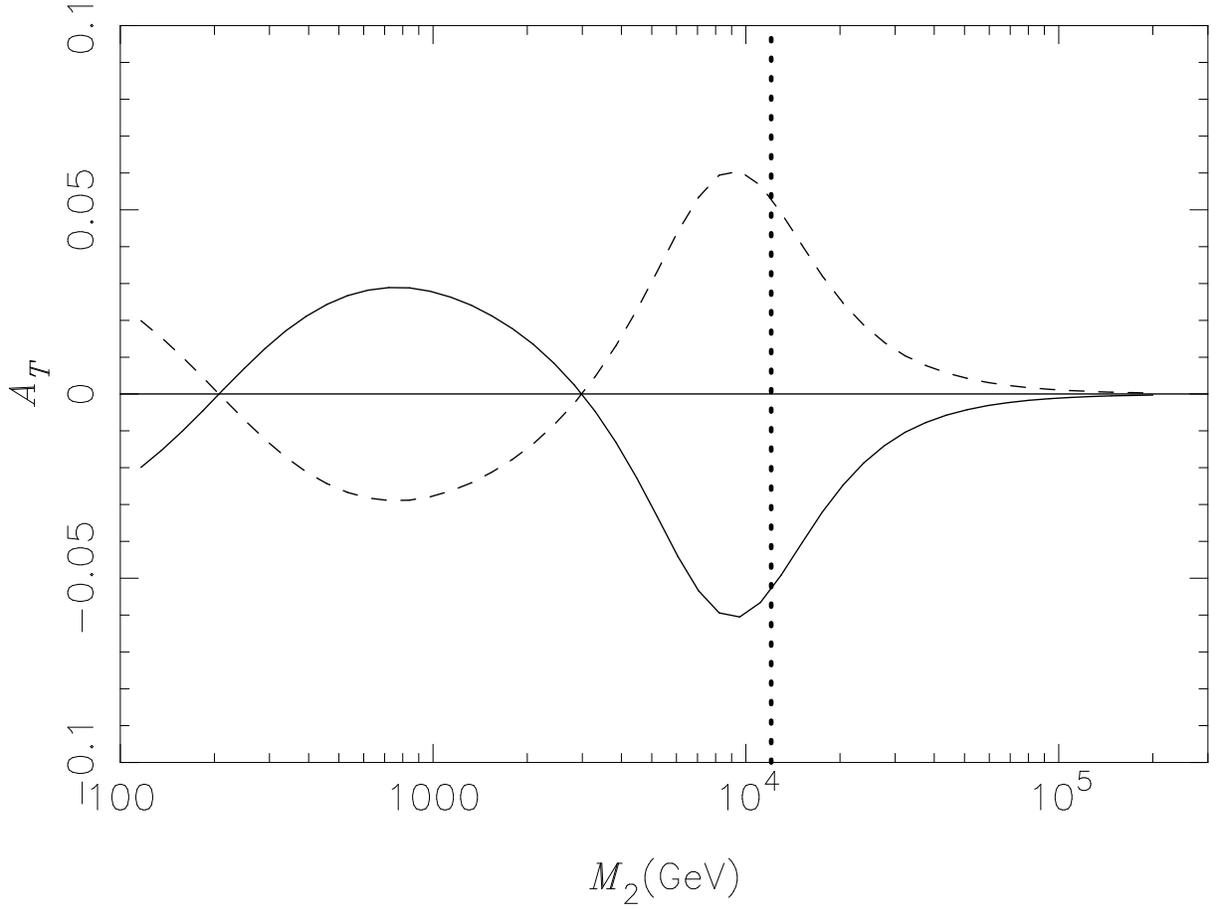}
\end{center}
\vskip 1.5cm
\caption{The T-asymmetry $A_T$  in Model II, for the decay $\tau
\to 3e$ as a function of $M_2$, keeping $M_1/M_2\!=\!0.5$ and
adjusting the mass parameters in order to obtain maximal branching
ratios [{\it i.e.} Fig.~2(b)]. The CP phase is taken to be
$\xi\!=\!\pi/4$ (solid line) and $\xi\!=\!3\pi/4$ (dashed line).
$Br(\tau \to 3 e)$ for $\xi\!=\!\pi/4$ or $3 \pi/4$ is lower
by about factor two in comparison to the case $\xi\!=\!0$
of Fig.~2(b).}
\label{figAT} \label{figATMII}
\end{figure}
\end{center}

Fig.~\ref{figATMII} shows the T-odd asymmetry $A_T$ of
Eq.~(\ref{AT}) for the same choices of mass matrix parameters as
in Fig.~\ref{figBrMII}(b), but for $\xi=\pi/4$ (solid line) and
$\xi=3\pi/4$ (dashed line) (maximal CP violation). When all four
heavy Majorana neutrinos are degenerate, there is no CP violation
\cite{BRV} and $A_T\!=\!0$. Also  $A_T\!=\!0$ if $\xi\!=\!0$ or $\pi/2$.
Notice that the maximal $Br$'s are reached for $\xi\!=\!0$, thus no CP
violation;
in the cases of Fig.~\ref{figATMII} ($\xi\!=\!\pi/4, 3 \pi/4$)
the values of $Br$'s are about one half of the corresponding ones
in Fig.~\ref{figBrMII}(b) ($\xi=0$).
The searches for maximal rates and for CP violation are in
this sense complementary in their most optimistic cases.

In the $\tau \to \mu$ processes, the maximal branching ratios in
Model II are suppressed by an additional factor of
$(s_L^{\nu_{\mu}})^2_{\rm max}/(s_L^{\nu_{e}})^2_{\rm max} =
0.002/0.005 \approx 0.4$ [cf.~Eq.~(\ref{s2s})]. The
approximate maximal branching ratios for the $\mu \to e$ LFV
processes in Model II are obtained from the corresponding
$\tau \to e$ branching ratios by multiplying them with
$(s_L^{\nu_{\mu}})^2_{\rm max}/(s_L^{\nu_{\tau}})^2_{\rm max} =
0.2$ and dividing by $Br(\tau \to \mu {\overline \nu}_{\mu}
\nu_{\tau}) \approx 0.174$. Thus, $Br(\mu \to e \gamma)
\stackrel{<}{\sim} 10^{-8}$ and $Br(\mu \to 3e)
\stackrel{<}{\sim} 10^{-9}$, which is above the present
experimental bounds $1.2 \times 10^{-11}$ and $1.0 \times 10^{-12}$.
Therefore, the maximizing conditions
(\ref{amodII})--(\ref{bcdmodII}) cannot be met in this case.

\subsection{Expected numbers of events}

The explicit numbers of expected events in the considered
processes depend on the way the $\tau$ leptons are produced and on
the luminosities involved. For example, the $\tau$ pairs could be
produced via $e^+ e^- \to \tau^+ \tau^-$ close to the production
threshold, or by sitting on a specific vector meson resonance $V$:
$e^+ e^- \to V \to \tau^+ \tau^-$. In this case,
$\sigma(e^+ e^- \to V)$ as a function of the CMS energy
$\sqrt{s}$ can be approximated as a Breit-Wigner function
\begin{equation}
\sigma(s; e^+ e^- \to V) = K
\frac{1}{\left[(\sqrt{s} - M_V)^2 + (\Gamma_V/2)^2 \right] } \ ,
\label{BWform1}
\end{equation}
where $M_V$ and $\Gamma_V$ are the mass and the total decay width
of the resonance. Constant $K$ in
Eq.~(\ref{BWform1}) can be fixed by invoking the narrow width
approximation (nwa) formula
\begin{equation}
\sigma_{\rm nwa}(s; e^+ e^- \to V) = \frac{12 \pi^2
\Gamma_{ee}(V)}{M_V} \delta(s - M^2_V) \ , \label{nwa}
\end{equation}
where $\Gamma_{ee}(V)$ is the partial decay width for $V \to
e^+e^-$. Integration of (\ref{nwa}) over the variable
$s$ gives $12 \pi^2 \Gamma_{ee}(V)/M_V$, fixing the
constant $K$ in (\ref{BWform1}): $K\!=\!3 \pi
\Gamma_{ee}(V) \Gamma_V /M^2_V$. The production cross section is
maximal on the top of the resonance $\sqrt{s}\!=\!M_V$:
\begin{equation}
\sigma(e^+ e^-\!\to\!V\!\to\!\tau^+ \tau^-)^{\rm max} =
\sigma(e^+ e^- \to V)^{\rm max} \times
\frac{\Gamma_{\tau \tau}(V)}{\Gamma_V}
\approx  12 \pi \frac{\Gamma_{ee}(V)}{\Gamma_V}
\frac{\Gamma_{\tau \tau}(V)}{\Gamma_V} \frac{1}{M^2_V} \ .
\label{sigmax}
\end{equation}
Multiplying this cross section by twice the branching ratio
$Br(\tau \to e\gamma (eee))$ we obtain the cross section for the
process $e^+ e^- \to V \to \tau^+ \tau^- \to e\gamma (eee) +
\tau$. These branching ratios are $\stackrel{<}{\sim}\!10^{-8}
(10^{-9})$ in Model II, as shown in Figs.~\ref{figBrMII}. For
example, if the resonance is taken to be $V = \Upsilon(1S)$
($M_V\!=\!9.46$ GeV; $\Gamma_{ee}(V)/\Gamma_V \approx \Gamma_{\tau
\tau}(V)/\Gamma_V \approx 0.025$), then $\sigma (e^+ e^- \to V \to
\tau^+ \tau^- \to e\gamma(eee) + \tau)$ would be about $2.$
($0.2$) fb. For a luminosity of $10~{\rm fb}^{-1}/{\rm yr}$, this
corresponds to 20 (2) events per year. Increased luminosities
would give correspondingly larger numbers of events.

\section{Conclusions and Summary}

We investigated and compared heavy Majorana neutrino effects on
lepton flavor violating (LFV) decay rates of tau leptons, in two
popular models: (I) the interfamily seesaw-type model realized in
the SM with right--handed neutrinos, and (II) the SM with
left--handed and right--handed neutral singlets. Further, we
calculated a T--odd asymmetry $A_T$ for $\tau \to 3 l^{\prime}$.
Model I is severely constrained in most of its parameter space by
the actual eV--scale experimental upper bound on the light
neutrino masses. It can give maximal LFV branching ratios $Br(\tau
\to \mu \gamma) \sim 10^{-9}$ and $Br(\tau \to 3 \mu) \sim
10^{-10}$ in a very restricted region of parameter space (where,
incidentally, the CP-violating asymmetry $A_T$ is zero), but
otherwise it gives branching ratios many orders of magnitude
smaller. On the other hand, in Model II the LFV branching ratios
can be larger over a wide range of parameter values, $Br(\tau \to
e \gamma) \sim 10^{-8}$ and $Br(\tau \to 3 e) \sim 10^{-9}$, and
$A_T$ can reach values up to $5\%$. The results in Model II are
insignificantly affected by the experimental bounds on the light
neutrino masses. Model II can predict large enough LFV branching
ratios ($Br$'s) to be tested with near future experiments.
The maximal LFV $Br$'s in Model II are obtained when $A_T\!=\!0$,
and are reduced by about a factor of two when $A_T\!\approx\!5 \%$.
Model II thus indicates a certain complementarity in the searches of the
LFV decay rates and of the associated T--violation.
%
\acknowledgments

The work of G.C. was supported by FONDECYT (Chile), Grants
No.~1010094 and No.~7010094. C.D. received support in part from
FONDECYT (Chile), Grant No.~8000017 and Fundacion Andes (Chile),
Grant C-13680/4. G.C. and C.D. also acknowledge partial support
from the program MECESUP FSM9901 of the Chilean Ministry of
Education.  The work of C.S.K. was supported in part by CHEP-SRC
Program and Grant No. 20015-111-02-2, R03-2001-00010 of the KOSEF,
in part by BK21 Program and Grant No. 2001-042-D00022 of the KRF,
and in part by Yonsei Research Fund, Project No. 2001-1-0057. The
work of J.D.K was supported by the Korea Research Foundation Grant
(2001-042-D00022). We wish to thank S.~Kovalenko and A.~Pilaftsis
for helpful discussions.

\begin{appendix}

\section[]{Approximate maximization of the branching ratios}
\setcounter{equation}{0}

The amplitude squared for the $l \to l^{\prime} \gamma$
LFV process is, according to Eq.~(\ref{llgamma}), approximately
proportional to
\begin{eqnarray}
|A|^2 & \propto & {\Big|}
\sum_{\ell=1}^{N_L} B^{\ast}_{l\ell} B_{l^{\prime} \ell}
\times f(0)
+ \sum_{h=N_L+1}^{N_L+{\tilde N}_R} B^{\ast}_{lh} B_{l^{\prime} h}
\times f(m_h^2) {\Big|}^2 \ ,
\label{A2f}
\end{eqnarray}
where the first sum runs over the light, practically massless,
intermediate neutrinos, and the second sum over the heavy
neutrinos with masses $m_h$ ($\sim M_1 \sim M_2$).
The function $f$ depends of the mass of the
exchanged neutrino; in the specific case,
it is the loop function $G_{\gamma}$ appearing in Eq.~(\ref{Gg}).
We now approximate the second sum by assuming that all
the heavy neutrinos eigenmasses $m_h$ are the same: $m_h = M$.
Then this, together with the unitarity of
the matrix $U$ (note: $B_{li} = U^{\ast}_{il}$), implies
\begin{eqnarray}
|A|^2 & \propto & {\Big|}
\sum_{h=N_L+1}^{N_L+{\tilde N}_R} U_{h2} U^{\ast}_{h1} {\Big|}^2
\times |f(M^2) - f(0)| \propto
{\Big|}
\sum_{h=N_L+1}^{N_L+{\tilde N}_R} U_{h2} U^{\ast}_{h1} {\Big|}^2
\ .
\label{A2s}
\end{eqnarray}
Here we denoted the flavor index $l$ of the heavier charged lepton
as $2$ and the index $l^{\prime}$ of the lighter one as $1$.

The amplitude squared for the
$l \to 3 l^{\prime}$ LFV process is,
according to Eqs.~(\ref{gammal})--(\ref{Boxl})
and (\ref{Gg})--(\ref{FBox}), more complicated. However,
in the leading order in $m_D m_M^{-1}$, and when there
is no CP violation (when matrix $U$ is real), it is
straightforward to show that $|A|^2$ is proportional
to the same kind of combination (\ref{A2f}). Thus,
the proportionality (\ref{A2s}) is approximately
satisfied also for the $l \to 3 l^{\prime}$ LFV process.

The above proportionality can be supplemented by
the Schwarz inequality
\begin{eqnarray}
|A|^2 & \propto & {\Big|}
\sum_{h} U_{h2} U^{\ast}_{h1} {\Big|}^2
\leq
\sum_{h} |U_{h2}|^2 \times \sum_{h^{\prime}} |U_{h^{\prime}1}|^2
\equiv (s_L^{\nu_2})^2 (s_L^{\nu_1})^2 \ .
\label{A2t}
\end{eqnarray}
The equality is achieved only when there is a proportionality:
\begin{equation}
\frac{U_{h2}}{U_{h1}}\Bigg{|}_{h=N_L+1} =
\frac{U_{h2}}{U_{h1}}\Bigg{|}_{h=N_L+2} =
\cdots = \frac{U_{h2}}{U_{h1}}\Bigg{|}_{h=N_L+{\tilde N}_R}
\label{Schweq}
\end{equation}

Thus, the approximate maximal value of $|A|^2$, and
thus of the LFV branching ratios, is achieved when
the values of the heavy--to--light mixing parameters
$(s_L^{\nu_2})^2$ and $(s_L^{\nu_1})^2$
are saturated according to the upper bounds
(\ref{s2s}) and, at the same time,
the mixing matrix $U$ elements in the
heavy--to--light sector satisfy the equalities (\ref{Schweq}).

In the seesaw Model I (with $N_L={\tilde N}_R=2$),
the mixing matrix $U$ elements in the
heavy--to--light sector are
$U_{h2} = (m_M^{-1} m_D^{\dagger})_{h^{\prime}2}$ and
$U_{h1} = (m_M^{-1} m_D^{\dagger})_{h^{\prime}1}$,
where $h^{\prime}\equiv h\!-\!2$.
In our specific case of maximal mixing
($a\!=\!c$, $b\!=\!-d$, $\delta_1\!=\!\delta_2\!=\!\pi/2$) we have:
$U_{31}=a/M_1$, $U_{41}=-i b/M_2$,
$U_{32}=-ia/M_1$, $U_{42}=-b/M_2$; the
equality (\ref{Schweq}) is fulfilled; and
$(s_L^{\nu_1})^2 = (s_L^{\nu_2})^2 = (a^2/M_1^2 + b^2/M_2^2)$.

In Model II (with $N_L=2$ and ${\tilde N}_R=4$),
with $\xi=0$, it can be shown, e.g. by using
{\em Mathematica}, that the equality (\ref{Schweq})
is satisfied when $ad=bc$. If in this case also
the values of $(s_L^{\nu_2})^2 = \sum_{h} |U_{h2}|^2$
and $(s_L^{\nu_1})^2 = \sum_{h} |U_{h1}|^2$
are saturated by the corresponding
upper bounds of Eq.~(\ref{s2s}),
then the approximate maximal branching LFV ratios
are achieved.

\end{appendix}

\end{document}